\documentclass[11pt,twoside]{article}
\usepackage[dvips]{graphicx}
\usepackage{asp2006}
\usepackage{epsf}
\usepackage{lscape}
\newcommand{\Msun}{M$_\odot$}
\newcommand{\kms}{km~s$^{-1}$~}
\def\deg{\ifmmode^\circ\else$^\circ$\fi}  
\def\arcsec{\ifmmode {'' }\else $'' $\fi}  
\def\arcmin{\ifmmode {' }\else $' $\fi}    

\begin{document}
\title{Fuel for Galaxy Disks}  
\author{Mary E. Putman$^{1}$, Jana Grcevich$^{1}$, J.E.G. Peek$^{2}$}  
\affil{$^{1}$University of Michigan, Dept. of Astronomy, 500 Church St, Ann Arbor, MI 48109-1042}   
\affil{$^{2}$Dept. of Astronomy, University of California, Berkeley, CA 94720}
\begin{abstract} 
Halo clouds have been found about the three largest galaxies
of the Local Group and in the halos of nearby spirals.  This suggests
they are a relatively generic feature of the galaxy evolution process and a
source of fuel for galaxy disks.  In this review, two main sources of disk star formation fuel, satellite material and clouds condensing from the hot
halo medium, are discussed
and their contribution to fueling the Galaxy quantified.  The origin of 
the halo gas of M31 and M33 is also discussed. \end{abstract}

\vspace{-0.4in}

\section{Introduction}

Beautiful stellar disks only come to be with the accretion of gaseous clouds of
star formation fuel.   There are several reasons why this fuel is thought to be gradually accreted from a galaxy's halo.   The first is
that halo gas exists around our Galaxy and other spiral galaxies (e.g., Oort 1970; Thilker et al. 2004; 
Oosterloo, Fraternali \& Sancisi 2007).   Halo gas therefore appears to be a relatively common phenomenon and the gas velocities indicate it will not
escape from the galaxy, but rather eventually fall towards the disk.  
 The second reason is the metallicity distribution of the long-lived stars in the Galactic disk,
 which indicates low metallicity fuel must be continually accreted, i.e., the G-dwarf problem.
The metallicity distribution of the G and K dwarfs in the solar neighborhood cannot be
reproduced with simple
closed box models, and the need for gaseous
inflow for the majority of the life of the disk has persisted with further observations
and increasingly complex chemical evolution models  (e.g., Larson 1972; Fenner \& Gibson 2003; 
Kotoneva et al. 2002; Chiappini, Matteucci \& Romano 2001; Magrini, Corbelli \& Galli 2007; Worthey et al. 2005).
A third reason to bring in star formation fuel from the halo 
is the lack of radial gaseous inflow observed in the disk of spiral galaxies (Wong et al. 2004).
Bringing the outer, relatively unenriched gas to the inner regions of the disk would
potentially provide a fresh source of star formation fuel.  

There are two main potential sources of gaseous fuel in a galaxy halo:  gas-rich satellites and 
condensed material from the hot diffuse halo.   
The latter may be a combination of material
left from the initial collapse of baryons into the dark matter halo, stripped material from
the satellites that was integrated into this hot halo, and some level of galactic
fountain material at low latitudes.   In this review, we discuss these two reservoirs of disk star 
formation fuel with a focus on the Milky Way, given this is where we have 
the most information on the satellite distribution and the low column density halo gas.  We will
also discuss the halo gas of M31 and M33 in the context of its origin and fate.

\begin{figure}[!ht]
\plotfiddle{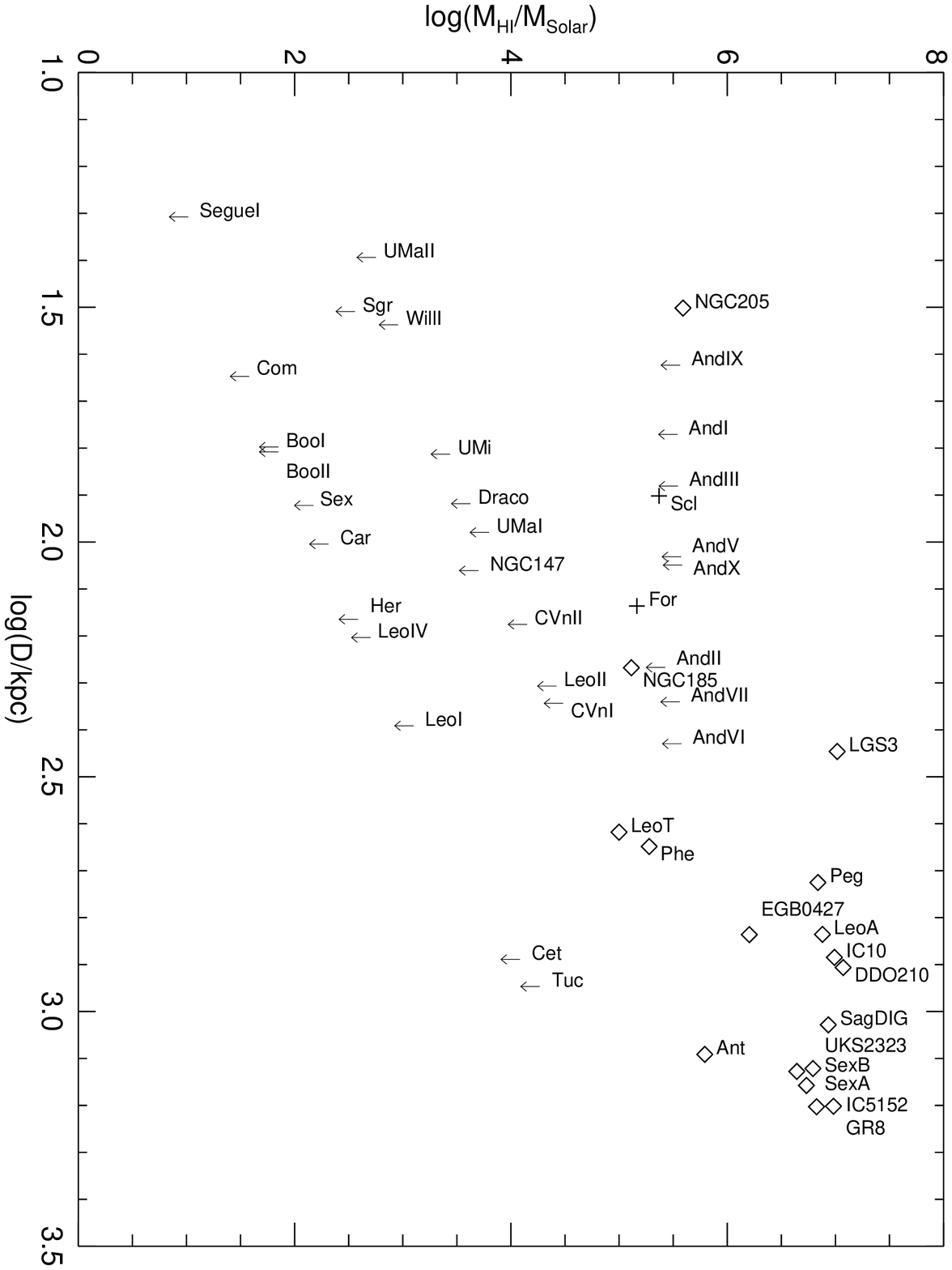}{2.9in}{90.0}{50.0}{50.0}{180.0}{-40.0}
\caption{The galactocentric distance to dwarf satellites of the Galaxy or M31 verses the HI mass or limit on the HI mass of the satellite (Grcevich \& Putman 2008).   HI detections are shown by the open diamonds and the majority are beyond 300 kpc from 
the Galaxy or M31.  The 
non-detections (or limits) are shown by downward arrows, and Sculptor
and Fornax are labeled as ambiguous detections as there are clouds in the
vicinity that may or may not be associated.\label{hidist}}
\end{figure}

\section{Accretion of Gas-Rich Satellites}

The number of satellites in the Local Group has increased substantially recently with the
results from the Sloan Digital Sky Survey (SDSS; e.g., Willman et al. 2005; Zucker et al.
2006; Belokurov et al. 2006) and surveys of the M31 environment (e.g., Ibata et al. 2007).  The new satellites have typical V-band luminosities of $10^{3-5}$ L$_\odot$ and total masses in the $10^{6-7}$ \Msun~range (Simon \& Geha 2007).   They are found
at a range of distances about both the Galaxy and M31.  Gas from satellites is generally the first thing to be stripped as it passes
through the diffuse halo medium of the larger galaxy.  The stripped gas will fuel, or has fueled, the Galaxy and M31.

\begin{figure}
\begin{center}
\includegraphics[height=2.3in]{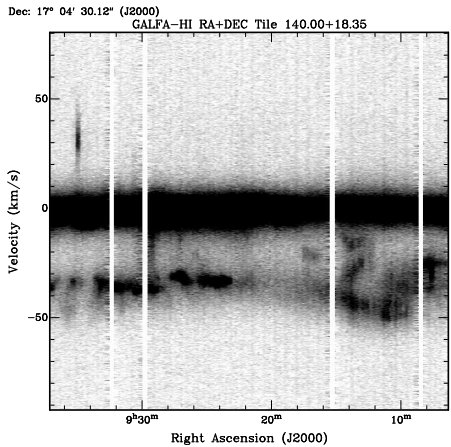}
\end{center}
\vspace{-0.08in}
\caption{A position-velocity plot of GALFA HI data showing the detection of the new SDSS 
gas-rich dwarf galaxy Leo T at $\sim35$ \kms and
the emission from the Galaxy.\label{leot}}
\end{figure}

Figure~\ref{hidist} shows the distribution of Galactic and M31 satellites in
distance and HI mass (Grcevich \& Putman 2008).  This plot is similar to Blitz \& Robishaw (2000), but is updated
to incorporate new HI data, optical velocities, and all of the known Local Group dwarf galaxies.   
The majority of the dwarf galaxies in the Local Group that still have neutral hydrogen are beyond 300 kpc 
from the Galaxy and M31\footnote{Excluding the Magellanic Clouds which together have a total mass $>10^{10}$\Msun.}.  In general those galaxies still with gas lie towards
the outskirts of the Local Group.   The gaseous detection of the new SDSS galaxy Leo T 
(Irwin et al. 2007) with the Galactic Arecibo L-Band Feed Array (GALFA) is shown in Figure~\ref{leot} (see also Ryan-Weber et al. 2008).
The majority of the galaxies that do not have
neutral hydrogen are within $\sim300$ kpc, with the exception of Tucana and Cetus.
There could be a SDSS sensitivity effect for why there are not more gas-less galaxies at large
radii (Koposov et al. 2007), but the result
of having all of the galaxies with HI beyond 300 kpc does not change.  The galaxies
without gas have most likely been subjected to ram pressure stripping (e.g., Mayer et al. 2006),
but halo densities high enough to effectively strip dwarf galaxies are not expected  
at radii of 150-300 kpc (see Grcevich \& Putman 2008 for details).

\vspace{-0.08in}

The amount of gas brought into the Milky Way from the accretion of
satellite galaxy material can be estimated by presuming that all of
its currently gas-less satellites had their fuel
assimilated into the Galactic disk.  Known stellar streams around the Milky Way can
also be included, though obviously we cannot account for the dwarfs that
have also had their stellar component completely assimilated in the disk.
The gas-less dwarf galaxies in Figure 1 most likely once had a similar
gas mass to the dwarf galaxies with gas, or $\sim
10^{5-7}$ \Msun.   The gas-less dwarf galaxies and 4 other destroyed
dwarfs traced by stellar streams (Newberg et al. 2007; Belokurov et al. 2007)
would therefore have contributed $\sim 10^{6-8}$ \Msun~ of HI to fuel the Galaxy.  
This value would increase with the inclusion of the satellites' molecular and ionized gas, 
but it is unlikely to increase by more than a factor of two
(e.g., Leroy et al. 2007).   Chemical evolution models of the Galactic disk typically
require an average infall rate of 1 \Msun/yr over the past 5-7 Gyr (Chiappini et al. 2001),
and therefore the dwarf galaxies appear to provide sufficient fuel for only $\sim10^{8}$ years.
It should be noted that the 1 \Msun/yr is an average rate of infall, and it
should be closer to half that value today (Chiappini, this proceedings).    
In any case, to accomplish the average of 1 \Msun/yr from only the accretion
of gas from small dwarf galaxies would require an average of 300 accretion events per Gyr.  
This is extremely high compared to the predictions of $\Lambda$CDM models (Zentner et al.
2005), and the number of satellites we currently see in the halo.   

There is a source of gaseous fuel associated with Galactic satellites 
that has not yet been discussed, the Magellanic
System.  This system is currently at $\sim55$ kpc and consists of the Small
and Large Magellanic Clouds, and the Magellanic Stream and Leading Arm that
are being stripped from the Clouds (Putman et al. 2003).  The Magellanic System will 
eventually contribute $\sim 10^9$ 
\Msun~ of fuel just in neutral hydrogen to the Galaxy.  
Galaxies such as the Magellanic Clouds and larger bodies
may have contributed to fueling the Galaxy and M31 in the past.  In this case, 
only an average of 1 accretion event per gigayear would be required; however
the stability of the disk would most likely be an issue (Stewart et al. 2007).

\section{Condensing Halo Clouds}

In the discussion of satellite galaxy material above, the focus was on
the accretion of gas directly in the cold form.  Another method a galaxy can 
obtain star formation fuel is by accreting clouds that have condensed within
the hot halo medium.   This hot halo is predicted to originate from the initial
collapse of baryons into the dark matter potential well, but currently it would
be a mixture of this material, stripped satellite gas that was integrated into
the halo before reaching the disk, and galactic fountain material, or hot gas rising
into the halo from the cumulative effect of supernovae in the disk.
Recent galaxy formation and evolution models have
the hot halo gas gradually cool over time, with halo clouds forming from thermal instabilities 
(Maller \& Bullock 2004; Lin \& Murray 2000).   This is in contrast to the original models
that have all of the hot halo gas cool and fall in at early times (White \& Rees 1978; White \& Frenk 1991). 
The recent models predict at $z=0$
the existence of both a hot halo medium that hides a large percentage
of a galaxy's baryons, and cool halo clouds that will gradually fuel the disk
(Maller \& Bullock 2004; Sommer-Larsen 2006; Kaufmann et al. 2006; Fukugita \& Peebles 2006; Connors et al. 2006).

\begin{figure}[h!]
\begin{center}
\vspace{-0.01in}
\includegraphics[height=1.5in]{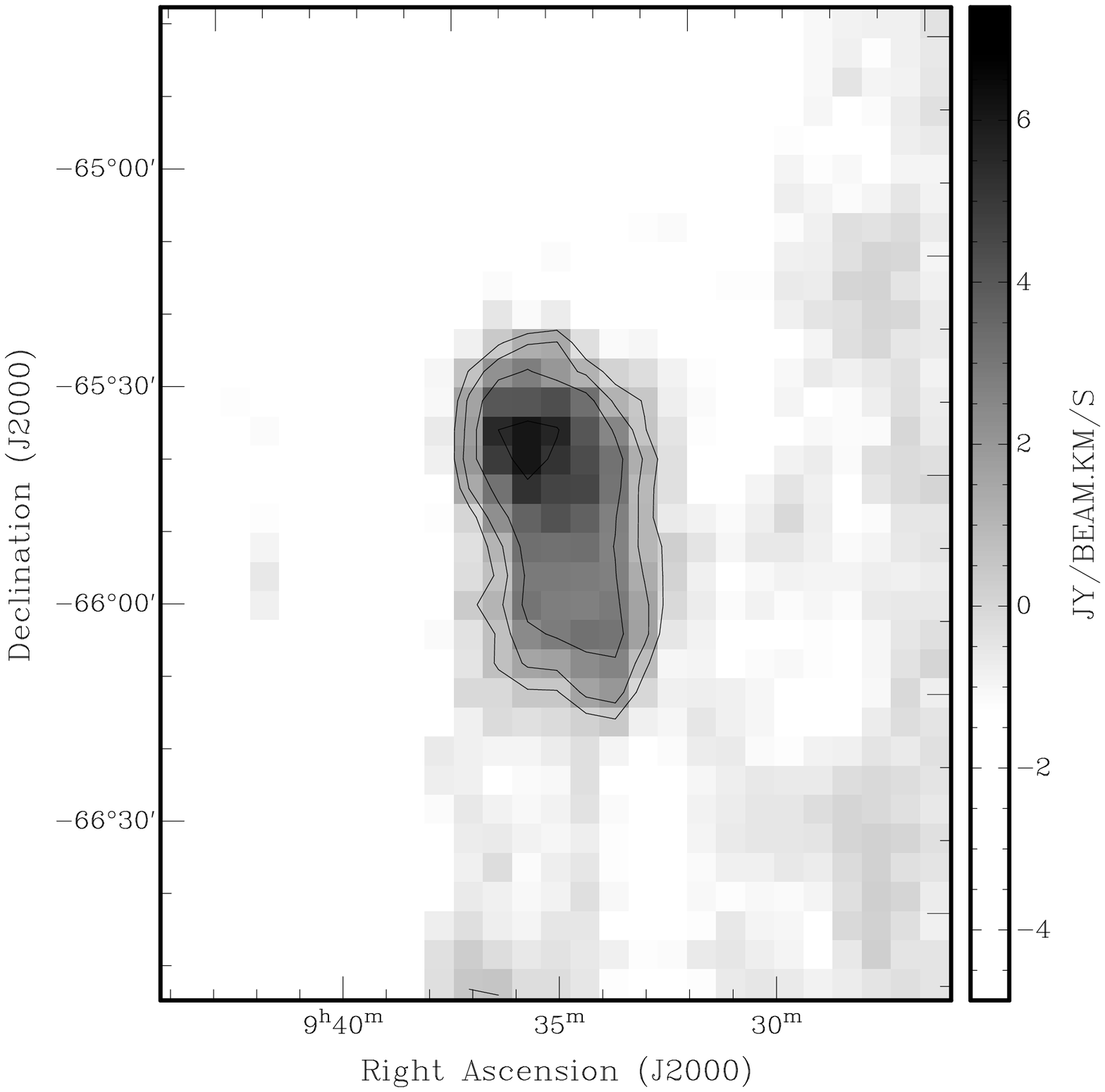}
\includegraphics[height=1.5in]{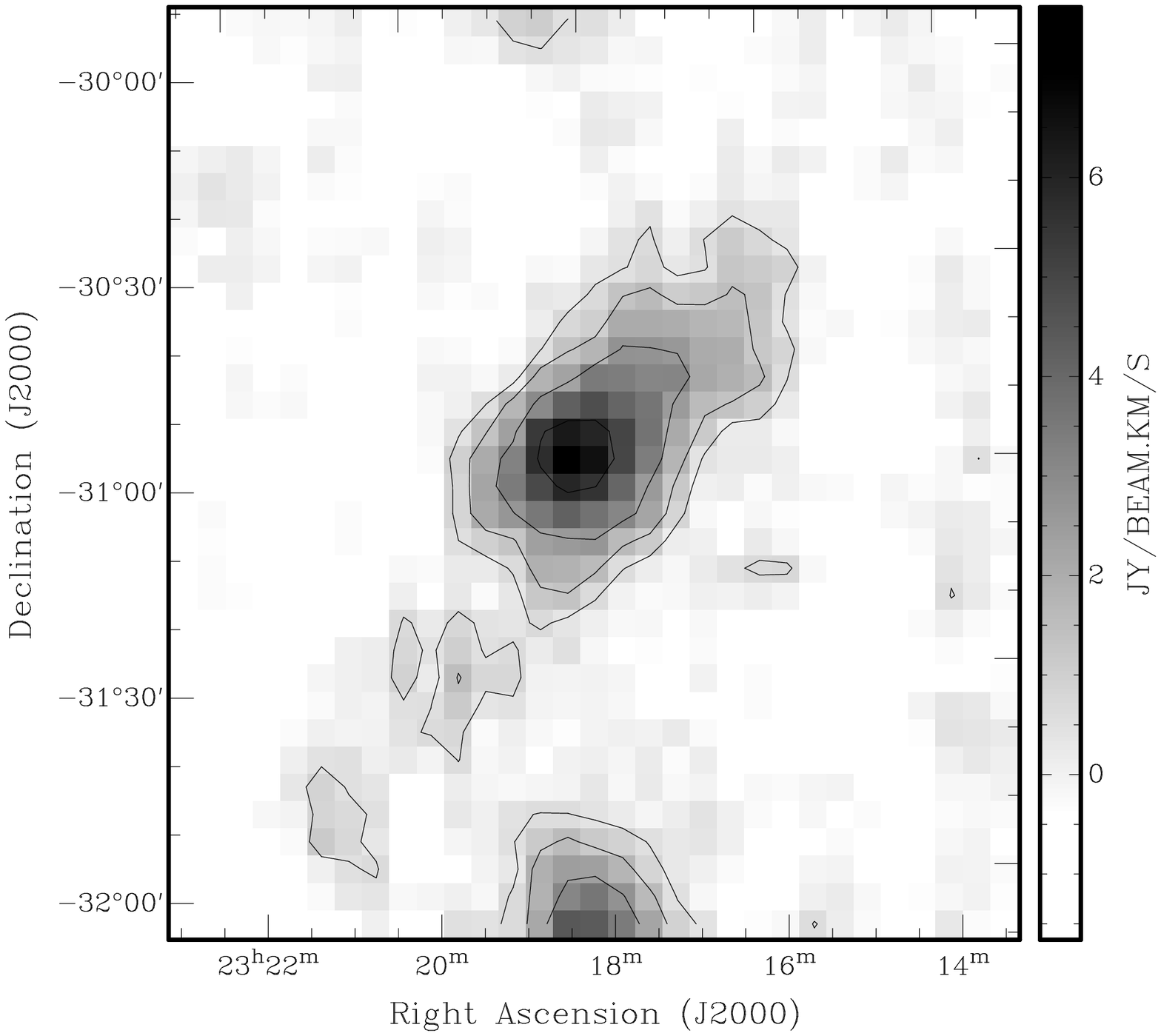}
\includegraphics[height=1.5in]{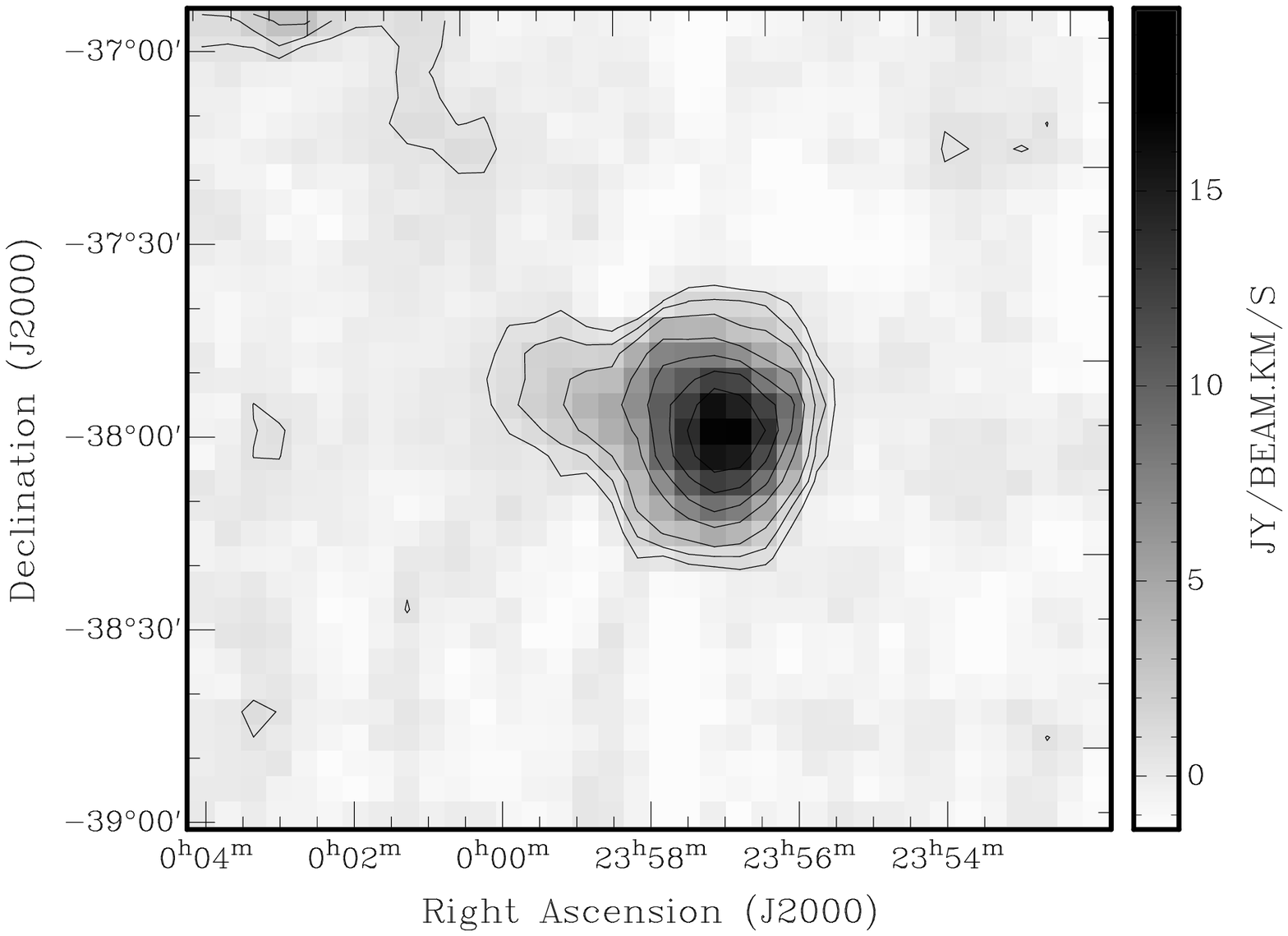}
\hspace{-0.20in}
\vspace{-0.20in}
\includegraphics[height=1.5in]{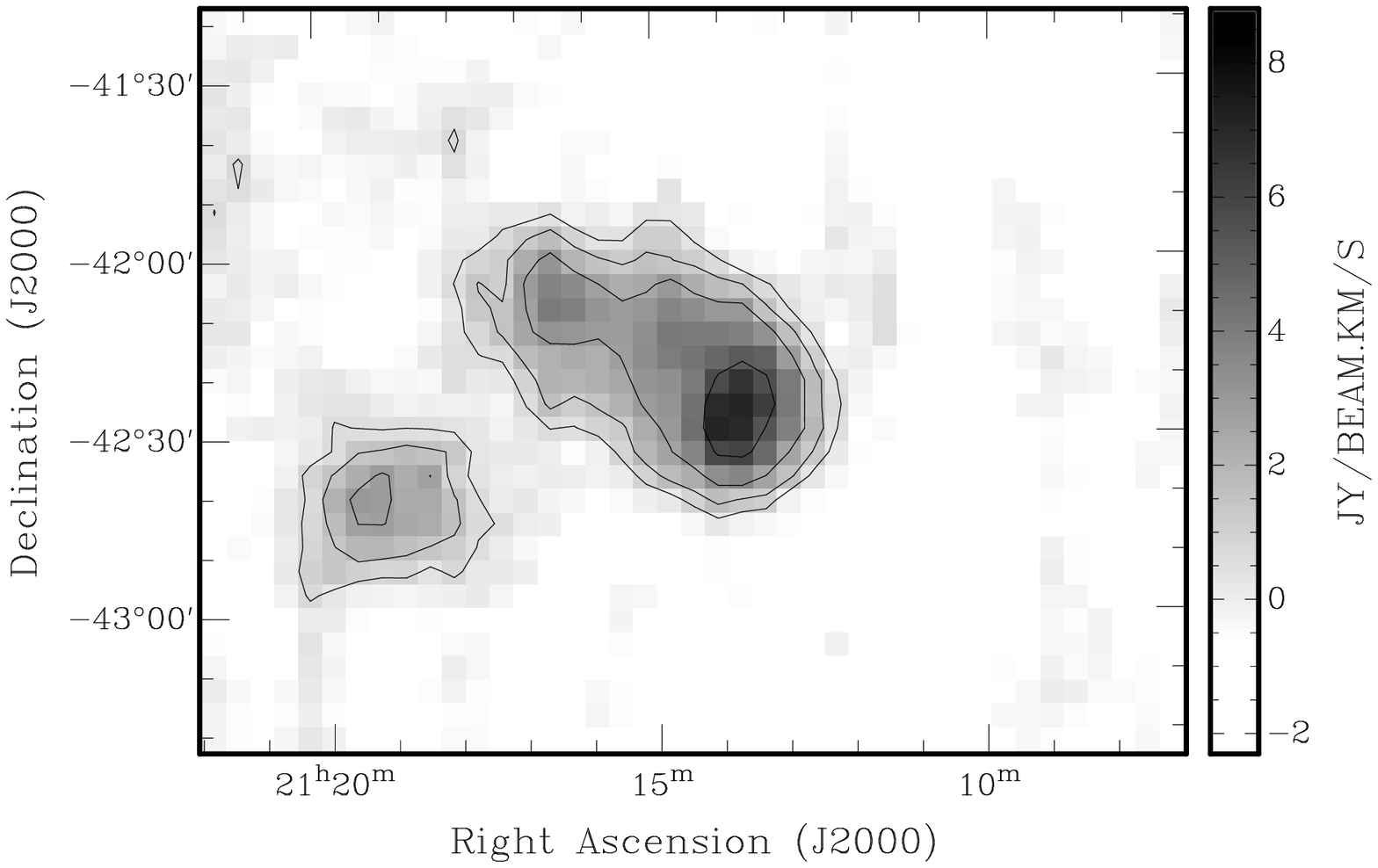}
\includegraphics[height=1.5in]{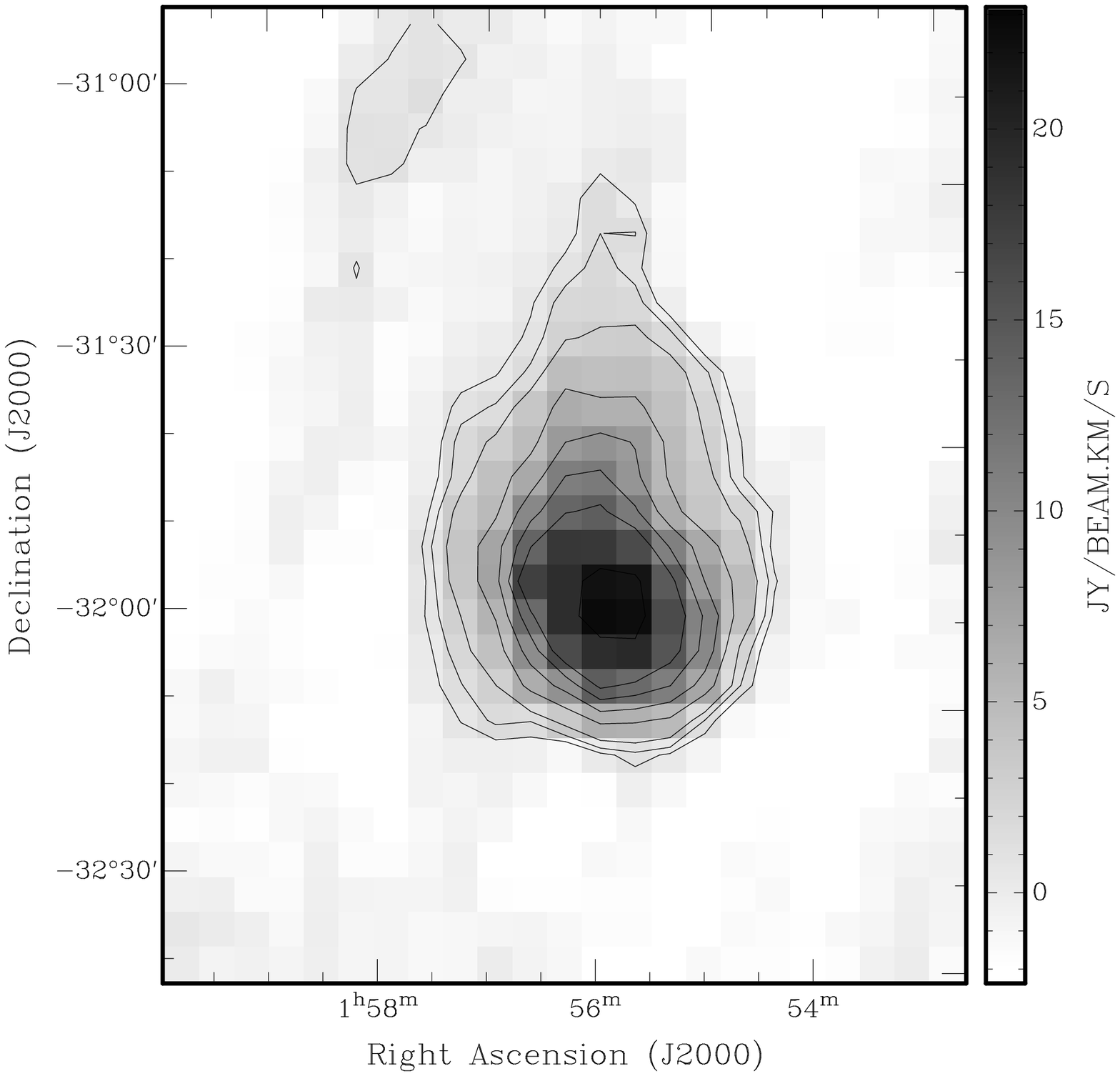}
\includegraphics[height=1.5in]{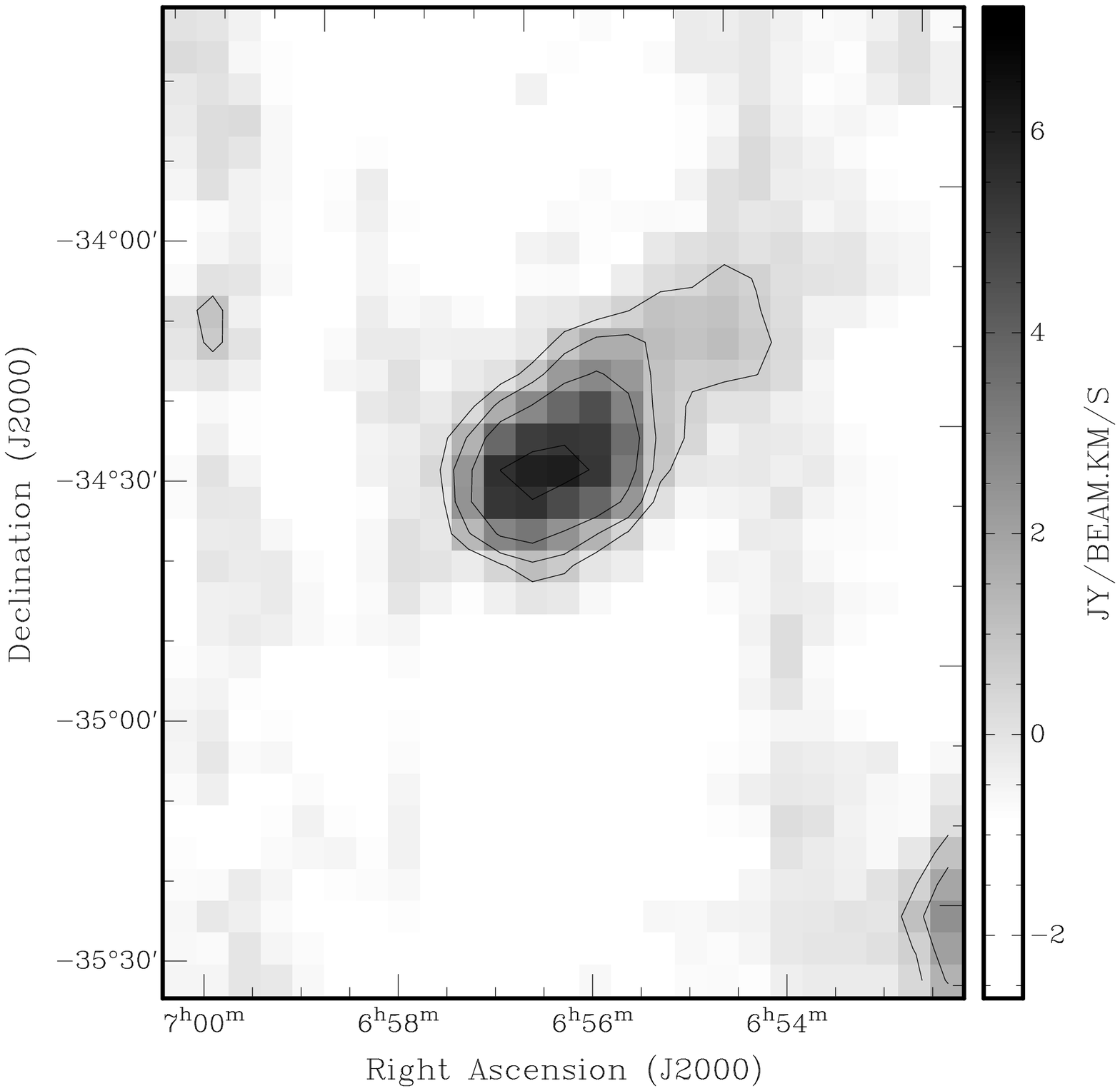}
\hspace{-0.20in}
\vspace{-0.20in}
\includegraphics[height=1.5in]{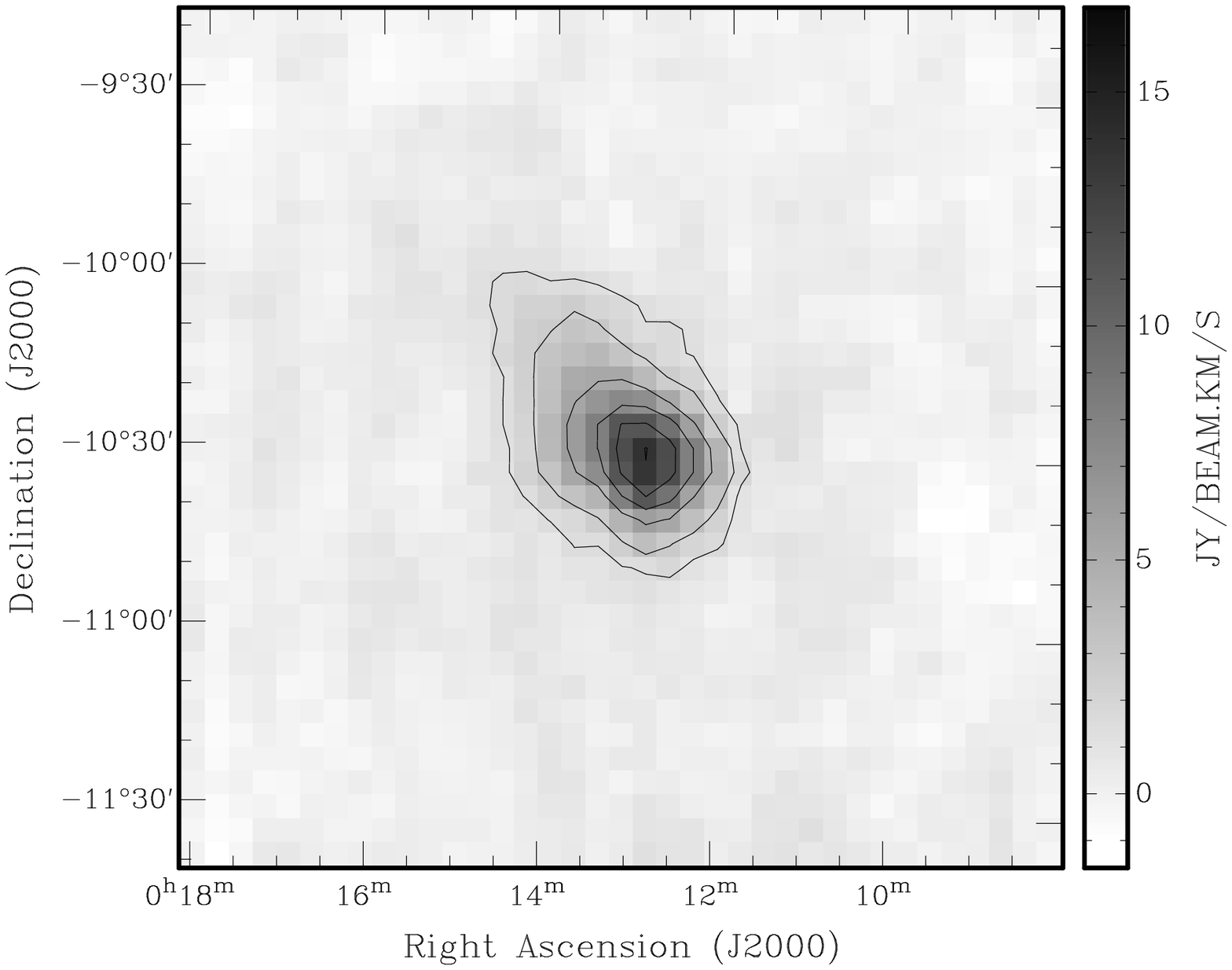}
\includegraphics[height=1.5in]{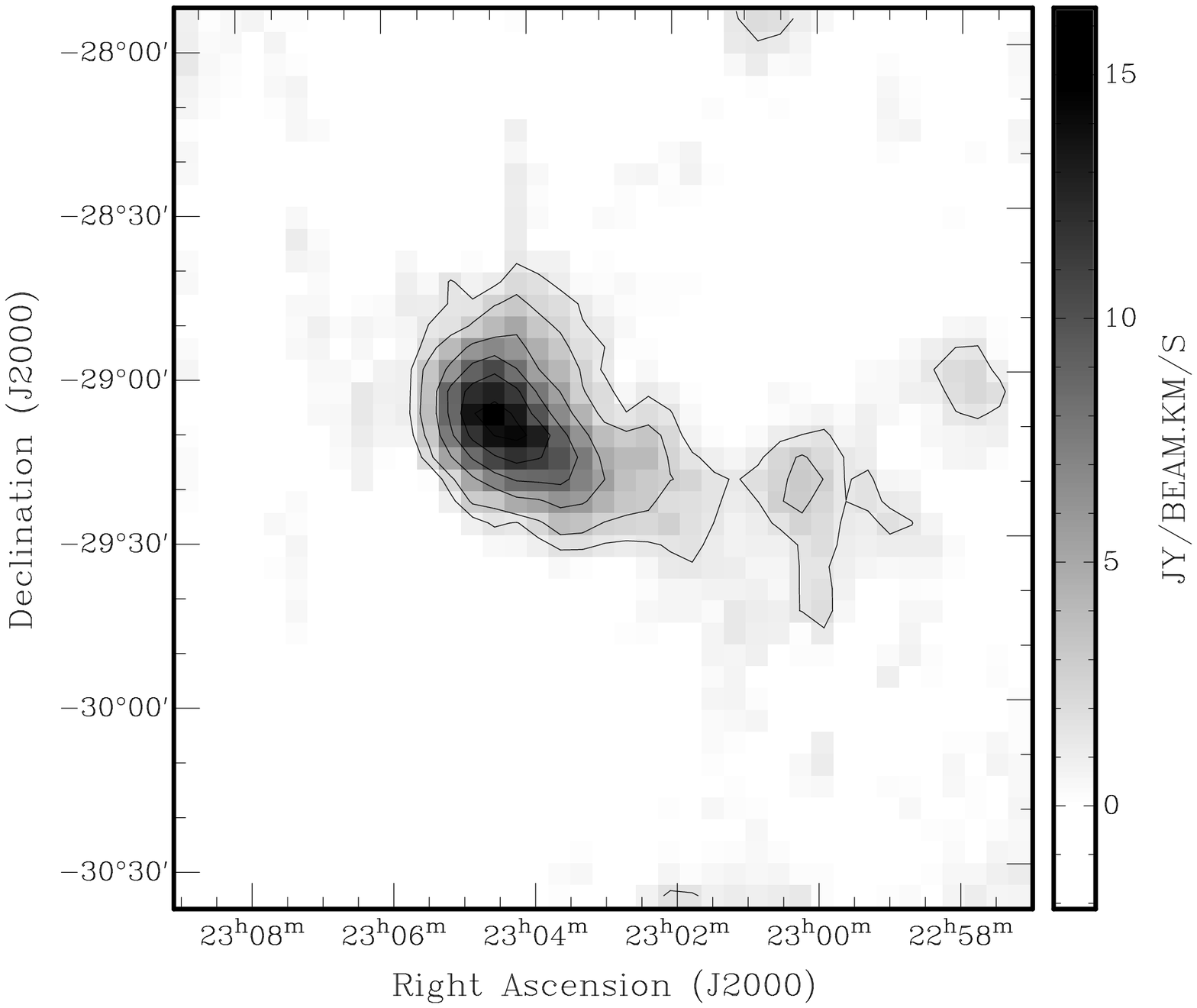}
\includegraphics[height=1.5in]{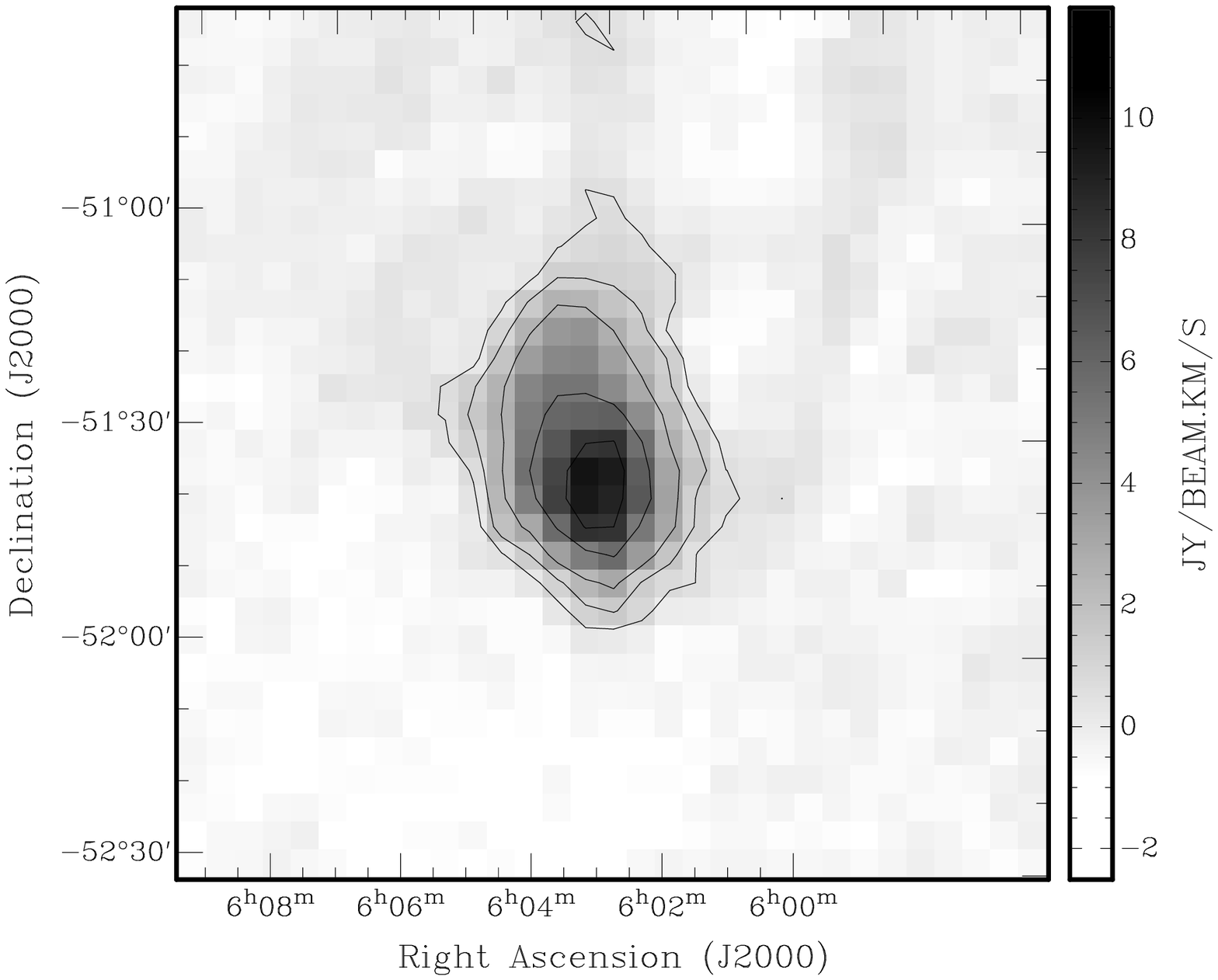}
\end{center}
\caption{A sample of the 166 head-tail clouds found in the HIPASS survey of HVCs (Putman et al. 2002).  The neutral hydrogen contours are 1.2 ($3\sigma$), 1.6, 2, 4, 8, 12, 16, and 20 $\times 10^{18}$ cm$^{-2}$.\label{ht}}
\end{figure}

Both cool halo clouds and a hot diffuse medium are found in the Galactic halo.
The cool halo clouds are the high-velocity clouds (HVCs) -- neutral hydrogen clouds with 
velocities that are inconsistent
with simple models of Galactic disk rotation (Wakker \& van Woerden 1997; Putman et al. 2002).  The hot diffuse halo medium is
largely detected indirectly through the structure of halo clouds (e.g., Peek et al. 2007)
and the detection of high-velocity O~VI absorption associated
with the interaction of the halo clouds and this medium (Sembach et al. 2003).
There are also constraints on the properties of the diffuse halo medium from x-ray absorption line detections (Fang et al. 2006; Wang et al. 2005; Williams et al. 2005).
Figure~\ref{ht} shows examples of head-tail clouds formed in the Galactic halo as HVCs move through the diffuse halo medium.   With Fabian Heitsch, we are simulating
this process to reproduce the structure of the observed head-tail clouds and probe
the properties of the diffuse halo medium.

In Peek, Putman \& Sommer-Larsen (2008) we examined the link between the
observed HVCs and the clouds produced in a high-resolution SPH cosmological simulation of galaxy
formation.  The simulation includes chemical evolution,
supernova feedback, and radiative cooling with an extragalactic UV
field (see Sommer-Larsen, Gotz \& Portinari 2003 for more details).
A resolution of $3 \times 10^5$ \Msun~for halo clouds was reached by 
finding a Milky Way (V$_{c} \sim 220$ km s$^{-1}$, M$\sim 10^{12}$ \Msun)
in a lower resolution simulation and then tracing it back and replacing one
particle with 12 at $z=0.4$, and then one particle with 8 particles $\sim0.2$ Gyr later.
The simulation was then run for another 0.5 Gyr and the halo
cloud population examined at several timesteps.

The simulated and observed halo clouds are found throughout
the Galactic sky, at distances below 60 kpc.  There has been substantial
recent progress on constraining the distances to the observed HVCs using
direct and indirect methods (e.g., Thom et al. 2007; Westmeier et al. 2007).
The simulated clouds have an average Galactic Standard of Rest velocity
of V$_{GSR} = -65$ to $-85$ \kms (depending on the location of the Sun
in the simulation), which is comparable to the observed HVC average 
velocity of V$_{GSR} = -85$ \kms.   The mass of the simulated clouds
can be converted to an observed flux with the known distance and 
by assuming a sensible ionized component ($< 50$\% ionized).  The simulated
cloud fluxes (individual and total) are also found to
be comparable to the observed HVCs. 
A large percentage of the flux from the simulated clouds comes from a few nearby
complexes, which is also the case for the Galactic HVCs (i.e. Complex C).
Peek et al. (2008) present the figures that support the comparisons discussed
in this paragraph.

\begin{figure}[!ht]
\plotfiddle{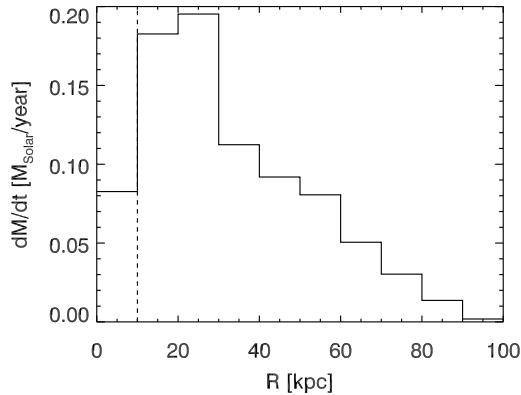}{2.0in}{90.0}{30.0}{30.0}{120.0}{-15.0}
\caption{The accretion rate of clouds onto subsequent galactic radii in the simulation.  The dotted line represents where the accretion rate can
no long be accurately determined because the clouds are merging
with the disk. See Peek et al. (2008) for details.  \label{infall}}
\end{figure}

The observed and simulated halo clouds show many similarities, indicating many
HVCs may represent clouds condensing from a hot halo medium.  The accretion
rate of these clouds can be determined from the simulation and is found to
be $\sim0.2$ \Msun/yr onto the galaxy's disk.  Figure~\ref{infall} shows the infall rate
of simulated clouds onto subsequent galactic radii.
The gradually increasing infall rate shows that clouds are created and/or grow at
all radii.  The convergence of the infall rate at 0.2 \Msun/yr is at $z=0$ and
this accretion rate is predicted to be much greater at higher redshifts consistent
with the higher star formation rate (Sommer-Larsen et al. 2003).

\section{Halo Gas of other Spiral Galaxies}

The previous sections focus on the accretion of gas onto the Galactic disk.
Up until recently the Milky Way was the only galaxy with a known halo
cloud population, with the exception of some detections of large halo features similar
to the Magellanic Stream ($\sim10^8$ \Msun) in nearby galaxies (e.g., Ryder et al. 2001; Hibbard et al. 2001).   
New receivers and larger
allocations of telescope time have allowed deep HI surveys of nearby spiral 
galaxies to be completed, and these surveys are beginning to indicate cold halo
clouds are a common property of spiral galaxies.  The Local Group spirals
are discussed here, but see also the deep HI observations of other nearby spirals
where additional halo clouds are observed
(e.g. Fraternali et al. 2001; Oosterloo et al. 2007; Heald et al., this proceedings; Battaglia et al 2006; Boomsma et al. 2005).

M31 is our closest spiral of similar size to the Milky Way, and was found to have numerous
halo clouds by Thilker et al. (2004).  The population of clouds detected is within 50 kpc 
of M31 and has a total HI mass of a few $\times 10^7$ \Msun. 
Westmeier et al. (2007) later showed there are no additional clouds
at larger radii.
Several of the M31 halo clouds are in the vicinity of the giant stellar stream (to the south of
M31) and NGC~205, while others appear to be completely isolated.   
The clouds in the vicinity of satellites may be remnants of their gaseous components, but they may
also represent clouds formed in instabilities in the hot halo medium created by the passage of the satellite.  
The isolated M31 halo clouds are most likely condensed clouds,
although they could also represent the remnants of clouds stripped 
from satellites at earlier times.   The velocities of the clouds relative to M31 and
the known satellites may cast some light on the most likely origin (e.g. see
figure 3 of Peek et al. 2008).
The total mass in M31's HI halo clouds
 is within a factor of three of the Galaxy's population if you adopt the distance constraint of
$< 60$ kpc for the Galactic halo clouds and exclude the Magellanic Stream (Putman 2006).

\begin{figure}[!ht]
\plotfiddle{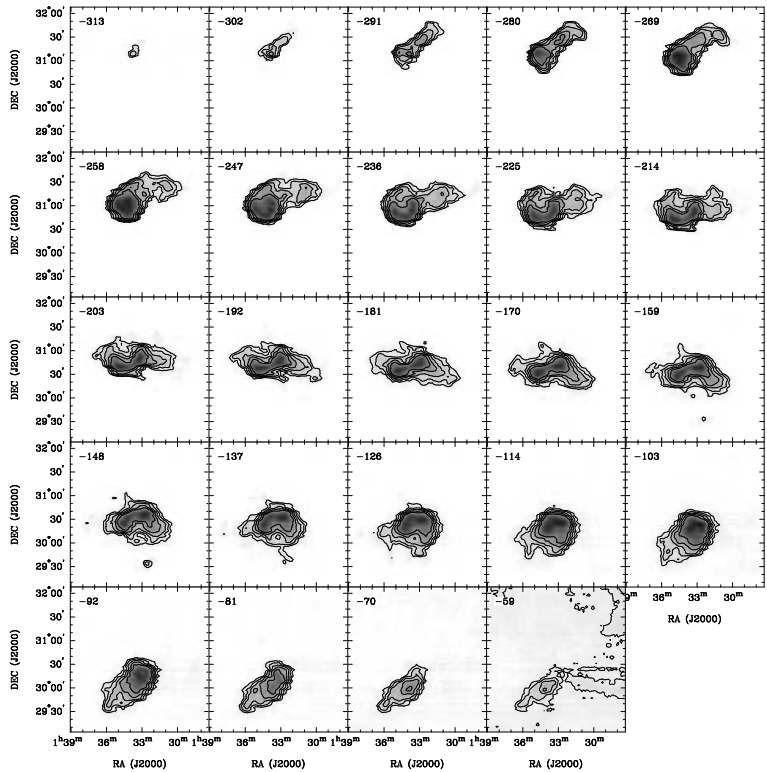}{3.9in}{-90.0}{52.0}{52.0}{-200.0}{300.0}
\caption{Channel maps of the HI in M33 with GALFA data.  The velocity is labeled in the upper left. Contours are 0.2, 0.4, 0.8, 1.6, 3.2 K.  The lowest
velocity channel shows some of the Galactic emission that overlaps with the emission from M33.  \label{m33chans}}
\end{figure}

M33 also has clouds in its halo. 
Figure~\ref{m33chans} shows the preliminary GALFA channel maps of
M33.
Two main halo features can be noted in this figure, though additional clouds at lower contour levels are evident in the data cube.  The first feature is most predominant at $\alpha, d = 1^{h}31^{m}, 31.5$\deg~ and -240 to -260 km s$^{-1}$, but continues beyond this velocity range and spatially wraps around to join the northern side of M33's warped disk.
The second halo feature is a small cloud with a filament to
the main galaxy at $\alpha, d = 1^{h}32.5^{m}, 29.5$\deg~ which is most evident at -148 \kms (see
also Westmeier et al. 2005).
The first halo cloud feature potentially has an interesting link to what may be a stellar
overdensity in the maps of Ibata et al. (2007); but given the limited coverage of
their current survey, it is unclear if this is actually a distinct stellar feature.
The potential continuation of this northern HI emission towards M31 
(Braun \& Thilker 2004) appears to be due to contamination by a cloud that merges with 
Galactic emission in the GALFA data.  The GALFA data extends up to $d = -34.5$\deg, and with its sensitivity ($3\sigma \sim 4 \times 10^{18}$ cm$^{-2}$ in the 11 \kms channels
shown in Figure~\ref{m33chans})
and resolution (3.5\arcmin~and 0.2 km s$^{-1}$) reveals the detailed kinematic and spatial structure of numerous low velocity
clouds that continue into the Galactic disk.
This link to Galactic emission would have been partially obscured in the Braun \& Thilker data with the 
smoothing and exclusion of examining Galactic emission.

The finding of halo clouds around M33 is somewhat surprising given M33 is
only $\sim$1/10 the mass of the Milky Way ($5\times10^{10}$ \Msun; Corbelli 2003).
Models do not generally expect a galaxy of this size to have a large amount of mass
locked up in its halo at $z=0$ (Maller \& Bullock 2004).
The clouds in the halo of M33 may represent the cold accretion mode
that is expected to dominate for galaxies of this mass (Keres et al. 2005), or the structures
may represent the gradual destruction of M33 by M31 which is now $\sim 200$ kpc
away.   The recent proper motion measurements of M33 by Brunthaler et al. (2005)
have been used by several authors to put constraints on the motion of M31 and its
impact on M33.
Loeb et al. (2005) rule out a large region in parameter space because of the lack
of disruption evident in the stellar component of M33.  van der Marel \&
Guhathakurta (2007) use the motion of M33, along with the motions of
other Local Group satellites, to calculate the most likely motion of 
M31.  The most likely transverse velocity for M31 has M33 on a tightly
bound orbit and signatures of disruption are therefore likely.
Muratov and Gnedin at the University of Michigan are completing more detailed
orbit calculations and finding similar results.
The HI features discussed here are most likely the beginning of this disruption process,
and the disruption is currently only evident in the extended gaseous component, similar
to the Magellanic Clouds.   It is possible M33 will eventually form giant HI features akin to those found in the
Magellanic System as it approaches perigalacticon.   Both of these
systems have a similar total HI mass ($1-2 \times 10^9$ \Msun) and will provide a substantial
amount of fuel to the galaxies.  

\section{Conclusions}
   
  There is growing evidence that gas clouds in galaxy halos are a common
  phenomenon and a natural part of the galaxy evolution process.  
  This halo gas is part of accreting satellites in some cases, but some of the clouds
  are consistent with a condensing cloud model.  
  The hot material the clouds
  condense out of at $z=0$ most likely consists of a mixture of materials
  that has been gradually enriched over time.   Since the medium is thought to be responsible for
  much of the stripping of dwarf galaxies (Mayer et al. 2006), and extends out to at least 55 kpc as evident from the observations
  of clouds in the Magellanic Stream (Stanimirovic et al. 2002; Putman et al. 2003; Sembach et al. 2003), it is unlikely to all originate from
  feedback from the Galaxy itself.  In fountain scenarios the hot gas
  generally rises to only $\sim10$ kpc (Booth \& Theuns 2007; de Avillez 2000). 
  One plausible scenario is the hot halo is created with the initial collapse
  of baryons into the dark matter potential well and is supplemented through
  time with fountain material and stripped gas from satellite galaxies.
  
  In terms of the Local Group galaxies, the three largest galaxies all show evidence
  for HI halo clouds with a total mass in clouds of $\sim 10^{7-8.5}$ \Msun.  This
  is not a large reservoir of gas in the context of the chemical evolution models,
  but with a continual new supply through incoming gas-rich satellites and condensing
  clouds it may be sufficient at $z=0$.  Presumably both of these sources were more 
  abundant at higher redshift.  The future may be bright for both the Milky Way
  and M31 in the context of future star formation fuel with the accretion of gas from
  the Magellanic System for the
  Milky Way and possibly M33's gas for M31.   There may also be additional Leo T's
  sitting at the outskirts of the Local Group that will serve as future Galactic and M31 morsels.
  The role of the diffuse halo of these galaxies will be further probed with observations
  and simulations of the head-tail
  clouds and new results from the Cosmic Origins Spectrograph.


\acknowledgements  M.E.P. thanks F. Fraternali, R. Sancisi, and J. Ostriker for useful discussions.  Many thanks to the 
Turn On GALFA Survey team, Snezana Stanimirovic, Kevin Douglas, Steven Gibson, Carl Heiles, and Eric Korpela for their help
with the GALFA data presented here.


\end{document}